\documentclass{article}

\usepackage{arxiv}

\usepackage[utf8]{inputenc} 
\usepackage[T1]{fontenc}    
\usepackage{hyperref}       
\usepackage{url}            
\usepackage{booktabs}       
\usepackage{amsfonts}       
\usepackage{nicefrac}       
\usepackage{microtype}      
\usepackage{lipsum}		
\usepackage{graphicx}
\usepackage{natbib}
\usepackage{doi}

\title{\textit{In-situ} observation of Alfv\'{e}n waves in ICME shock-sheath indicates existence of Alfv\'{e}nic turbulence}

\author{Anil Raghav$^{1*}$, Zubair Shaikh$^2$, Omkar Dhamane$^1$, Kalpesh Ghag$^1$, Prathmesh Tari$^1$,  Utsav Panchal$^1$ \\
\\
	$^1$Department of Physics, University of Mumbai, Mumbai, India\\
	$^2$Indian Institute of geomagnetism, Panvel, Navi Mumbai, India\\
	\texttt{*anil.raghav@physics.mu.ac.in} \\
}

\renewcommand{\shorttitle}{Raghav et. al. }

\hypersetup{
pdftitle={A template for the arxiv style},
pdfsubject={q-bio.NC, q-bio.QM},
pdfauthor={David S.~Hippocampus, Elias D.~Striatum},
pdfkeywords={First keyword, Second keyword, More},
}

\renewcommand{\shorttitle}{Raghav et. al. }

\hypersetup{
pdftitle={A template for the arxiv style},
pdfsubject={q-bio.NC, q-bio.QM},
pdfauthor={David S.~Hippocampus, Elias D.~Striatum},
pdfkeywords={First keyword, Second keyword, More},
}

\begin{document}
\maketitle

\begin{abstract}
The dynamic evolution of coronal mass ejection (CME) in interplanetary space generates highly turbulent, compressed, and heated shock-sheath. This region furnishes a unique environment to study the turbulent fluctuations at the small scales and serve an opportunity for unfolding the physical mechanisms by which the turbulence is dissipated and plasma is heated. How does the turbulence in the magnetized plasma control the energy transport process in space and astrophysical plasmas is an attractive and challenging open problem of the 21st century. For this, the literature   discusses three types of magnetohydrodynamics (MHD) waves/ fluctuations in magnetized plasma as the magnetosonic (fast), Alfv\'{e}nic  (intermediate), and sonic (slow). The magnetosonic type is most common in the interplanetary medium. However, Alfv\'{e}nic waves/fluctuations have  not been identified to date in the ICME sheath. The steepening of the Alfv\'{e}n wave can form a rotational discontinuity that leads to an  Alfv\'enic shock. But, the questions were raised on their existence based on the theoretical ground. Here, we demonstrate the observable in-situ  evidence of Alfv\'{e}n waves inside turbulent shock-sheath at 1 AU using three different methods desciribed in the literature. We also estimate  Els\"asser variables, normalized cross helicity, normalized residual energy and which indicate  outward flow of  Alfv\'en waves. Power spectrum analysis of IMF indicates the existence of Alfv\'{e}nic turbulence in ICME shock-sheath. The study has strong implications in the domain of interplanetary space plasma, its interaction with planetary plasma, and astrophysical plasma.   
\end{abstract}

\section{INTRODUCTION }
\label{sec:int}

 Coronal mass ejection (CME) is a huge cloud of solar plasma ( mass $\sim 3.2 \times 10^{14}~g$, kinetic energy $\sim 2.0 \times 10^{29}~erg$) submersed in magnetic field lines that are blown away from the Sun which propagates and expands into the interplanetary medium \citep{vourlidas2010comprehensive,howard2011coronal}. Their studies are of paramount importance given their natural hazardous effects on humans and the technology in space and ground  \citep{schrijver2010heliophysics,moldwin2008introduction,schwenn2006space}.
The propagation speed of CMEs is often higher than the ambient solar wind which causes the formation of fast, collision-less shocks ahead of CMEs \citep{kennel1985quarter}. These shocks cause heating and compression of the upstream (anti-sunward side) slow solar wind plasma, forming turbulent sheaths between the shocks and the leading edge of the CMEs  \citep{sonett1963distant,kennel1985quarter,zurbuchen2006situ,jian2006properties,echer2011interplanetary,richardson2011galactic,kilpua2017coronal}. 

The shock and sheath are responsible mostly for (i)  acceleration  of  solar  energetic  particles  \citep{tsurutani1985acceleration,manchester2005coronal,gosling1983ion,giacalone1994ion,zank2000particle,verkhoglyadova2015theoretical,zank2007particle,li2003energetic}, (ii)  significant geomagnetic activity \citep{tsurutani1988origin,shen2018shock,oliveira2014impact,oliveira2015impact,lugaz2016factors}, (iii)  Forbush decrease phenomena \citep{raghav2014quantitative,raghav2017forbush, raghav2020exploring,bhaskar2016role,shaikh2017presence,shaikh2018identification,shaikh2020evolution}, (iv) accelerate pickup ions \citep{giacalone1995simulations,gloeckler1994acceleration,zank1996interstellar}, and 
(v) auroral lightning \citep{baker2016resource} etc. Besides this, the shock initiates a magnetosonic wave in the magnetosphere and associated electric field accelerates electrons to MeV energies \citep{foster2015shock,kanekal2016prompt}. 

Recent theoretical \citep{zank2014particle,zank2015diffusive,li2003energetic,le2015kinetic,le2016combining,le2018self} and observational \citep{khabarova2015small,khabarova2016small,khabarova2017observational,zhao2018unusual,zhao2019particle, shaikh2018identification,shaikh2019coexistence,shaikh2020evolution,shaikh2020comparative,raghav2020pancaking} studies clearly indicate the local generation of quasi-2D structure \citep{zank2017theory,adhikari2017transport}, flux ropes \citep{shaikh2017presence} or magnetic islands in sheath region, and they may responsible for the acceleration of charged particles. 
Recently, the loss of electron flux from the radiation belt has been observed during the shock-sheath encounter with Earth's magnetosphere \citep{hietala2014depleting,kilpua2015unraveling,kilpua2017coronal}. This may be caused due to an increase in ultra-low frequency (ULF) wave power and dynamic pressure which is further responsible for pitch angle scattering and radial diffusion of the electron flux. The precipitated high energy electron flux from the radiation belt is used as a key parameter in climate models and the understanding of atmospheric chemistry and associated climatological effects \citep{verronen2011first,andersson2014missing,mironova2015energetic}. In addition to this, the other planets and their atmospheres are highly affected by the shock-sheath of CME, for example, in the case of Mars, loss of the ions flux ($>~9~amu$) is observed which might be caused by its high dynamic pressure \citep{jakosky2015maven}.

CME induced shock-sheath provides a unique opportunity to investigate the nature of plasma turbulence, plasma energy/fluctuation-dissipation, and plasma heating process.  The plasma turbulence demonstrates the features such as Alfv\'{e}n waves, Whistler waves, ion cyclotron waves, or ion Bernstein waves, etc \citep{krishan2004magnetic,gary2009short,schekochihin2009astrophysical,shaikh2010inhomogeneous,he2011possible,sahraoui2012new,salem2012identification}. In fact, sometimes plasma fluctuations do not exhibit any wave-like configuration at all but resemble nonlinear structures such as current sheets \citep{sundkvist2007dissipation,osman2010evidence}.
Various studies related to the nature of turbulence and generation of waves in the shock-sheath region have reported in the recent past. \cite{liu2006plasma} observed the mirror mode wave within the shock-sheath region. \cite{kilpua2013magnetic}
observed that the power of ultra-low frequency fluctuations (in the dynamic pressure and magnetic field) peaks close to the shock-front and sheath-magnetic cloud boundary. Furthermore, a large-amplitude magnetic field fluctuation, as well as intense irregular ULF fluctuations and regular high-frequency wave activity is also observed in the downstream of CME shocks \citep{kataoka2005downstream,kajdivc2012waves,goncharov2014upstream}. Moreover, Whistler waves associated with weak interplanetary shocks are also observed \citep{ramirez2012whistler}. 

The solar wind is predominantly associated with turbulent plasma \citep{bruno2005solar,bruno2013solar}, which contributes in acceleration of the solar wind  \citep{verdini2009turbulence,lionello2014validating}, solar wind heating \citep{freeman1988estimates,usmanov2011solar,adhikari2015transport}, and the scattering of the solar energetic particles \citep{li2003energetic,zank2007particle}. The turbulence of the solar wind plasma increases due to interplanetary shock \citep{burlaga1971nature,richter1985collisionless,jian2011comparing}. 
The particle acceleration rate is controlled by the shock strength, the turbulence level, the magnetic field strength, and the shock obliquity \citep{zank2000particle,zank2006particle}. It has been noted that the turbulence behind quasi-perpendicular shocks is more sporadic than that behind quasi-parallel shocks \citep{macek2015themis}. Note that, for quasi-perpendicular shocks, the cross-field currents are strong, produces significant levels of downstream plasma wave turbulence. Also, the shock steepening and the structure of shocks highly depends on the properties of the associated turbulence \citep{adhikari2016interaction}. Moreover, \citet{zank2015diffusive} demonstrated that the shock downstream turbulent, including vortical turbulence and Alfv\'enic like fluctuations is generated by the impact of upstream Alfv\'enic fluctuation disturbances. Recently,  \citet{zank2018theory} and \citet{adhikari2016interaction} studied the interaction between turbulence and termination shock and showed that quasi-two-dimensional turbulence dominates and slab-like turbulence plays a secondary role in the downstream of the shock wave. 

Besides this, various studies investigated Alfv\'enic fluctuations  and Alfv\'en waves  inside an ICME from $1 ~AU$ to $5.4 ~AU$ \citep{li2017evolution,marsch2009proton,yao2010identification,haoming2012alfvenic,LI_2013,li2016plasma,gosling2010torsional,raghav2018first} and calculated their contribution to local plasma heating\citep{li2017evolution}. ICME driven shocks are faster, stronger and show a larger distribution of shock parameters as compare to stream interaction shock \citep{kilpua2015properties}. Therefore, sheath plasma is expected to be highly compressed, hence pure Alfv\'en wave is not at all expected. Moreover, the turbulent nature of the sheath suggests the existence of Alfv\'enic fluctuations \citep{kataoka2005downstream} but explicitly not found to date. Besides, the literature indicates a theoretical debate on the existence of Alfv\'enic shock existence \citep{wu1987mhd,kunkel1966plasma,taniuti1964non} which may be expected during the steepening of Alfv\'en waves. Therefore, it is necessary to study the characteristics of Alfv\'en wave/fluctuations inside the highly turbulent sheath. Here we demonstrate the unambiguous {in-situ} evidence of Alfv\'{e}n wave suggests nature of Alfv\'enic turbulence within the shock-sheath region of CME.
\section{Event details}
\begin{figure*}
	\begin{center}		
		\includegraphics[width=\textwidth]{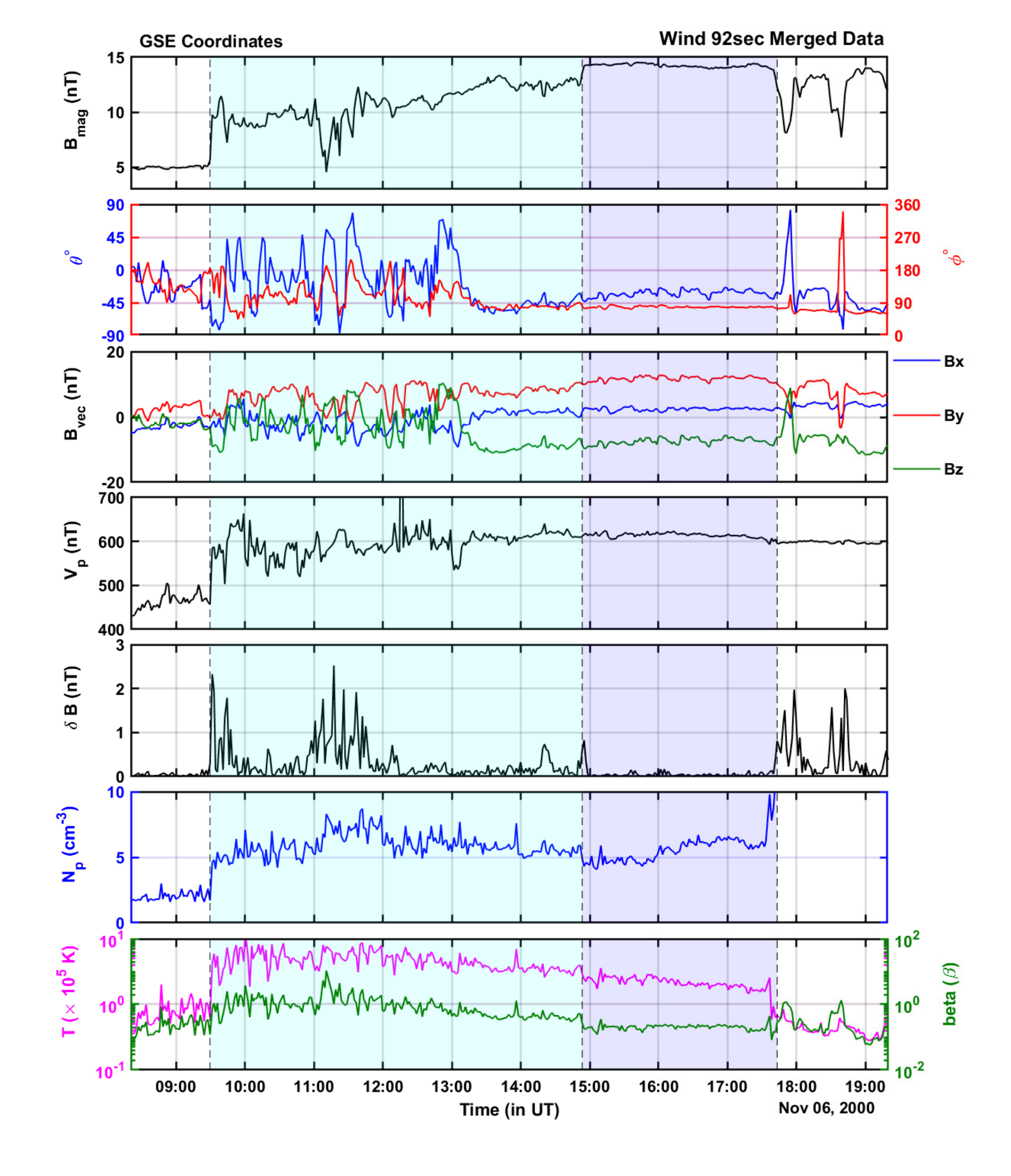}
	    \caption{The CME observed by the Wind spacecraft on $06^{th} ~ November, ~ 2000$ with time cadence of 92 sec. The top to bottom panels represents different interplanetary parameters such as: total interplanetary field strength IMF $B_{mag}$,elevation ($\theta^\circ = \arccos(\frac{-B_z}{B}) - 90^\circ$) $\&$  azimuth ($\phi^\circ = \arctan (\frac{-B_y}{-B_x}) + 180^\circ$)  angle,  IMF vectors i.e. $B_{vec}$, Plasma velocity ($V_p$), absolute value of IMF fluctuation i.e. $\delta B_i = \frac{B_{i+1} - B_{i-1}}{2}$,  Proton density ($N_p$), and  Temperature ($T_p$) $\&$ plasma beta ($\beta$) respectively. The shaded regions represents the shock-sheath of CME (cyan) and its magnetic cloud (purple). All observations are in GSE coordinate system. 
	     }
		\label{fig:IP}
	\end{center}
\end{figure*}

The shock-sheath under investigation is engendered by a CME which crossed the WIND and ACE spacecraft on $06^{th}$ November $2000$. Figure \ref{fig:IP} demonstrates the temporal variations of various in-situ plasma parameters and the interplanetary magnetic field (IMF) measured by the Wind spacecraft (The ACE spacecraft measurements are also studied, however not presented here). The commencement of the shock at spacecraft is identified as a sudden enhancement in the total IMF ($B_{mag}$), plasma beta ($\beta$), plasma density ($N_p$), plasma temperature ($T_p$), and  plasma speed ($V_p$), it is indicated by the first vertical black dashed line. In general, Rankine-Hugoniot condition is used to confirm the shock. The same condition is employed to the shock events observed by Wind spacecraft and its characteristics are given online at \url{https://www.cfa.harvard.edu/shocks/ac_master_data/00076/ac_master_00076.html}.  The shock is followed by large fluctuations in IMF (See $\delta B$ variations in the fifth panel of Figure \ref{fig:IP}) with enhanced magnetic field strength; high $N_p$, \& $T_p$  which is manifested as a shock-sheath region. The second shaded region shows the least fluctuations in $B_{mag}$ and its components, the slow variation in $\theta$ and $\phi$, the slow steady trend in $V_p$, and low $\beta$. This indicates the presence of a ICME magnetic cloud region \citep{zurbuchen2006situ}. The studied event is also listed in ICME catalogs available online e.g., \url{https://wind.nasa.gov/
list_plot_Wind/20001106_311_wind.png},
\url{http://www.srl.caltech.edu/ACE/ASC/DATA/level3/icmetable2.htm},
and \url{http://space.ustc.edu.cn/dreams/wind_icmes/web/png/WIND_20001106_223050_20001107_174216.png}.

\section{Alfv\'{e}n wave identification}

In literature, an Alfv\'en velocity is defined as:
\begin{equation}
V_A = \frac{B}{\sqrt{\mu_0 \rho}}
\end{equation}
where B is a magnetic field and $\rho$ is proton  density.  A typical Wal\'{e}n test is employed to confirm the presence of the Alfv\'{e}n wave in the solar wind. The Wal\'{e}n relation is described as \citep{walen1944theory,hudson1971rotational,yang2013alfven,yang2016observational,raghav2018first,raghav2018does}:
\begin{equation}
\Delta V = R_w \Delta V_A
\end{equation}
where, $R_w$ is the Wal\'en slope, $\Delta V$ and $\Delta V_A$ are the fluctuations in solar wind speed and fluctuations in Alfv\'en speed (magnetic field) respectively. The presence of Alfv\'en wave/ variations in the solar wind is suggested by a high correlation between the corresponding components of $\Delta V$ and $\Delta V_A$ as well as $R_w = \pm1$. The estimation of the correct background magnetic field and solar wind speed is essential to deduce fluctuations in their respective components. In this study, we confirmed the presence of Alfv\'en waves /fluctuations in the sheath region using three different methods as follow;  
\subsection{Method 1}
\begin{figure}
	\includegraphics[width= \textwidth]{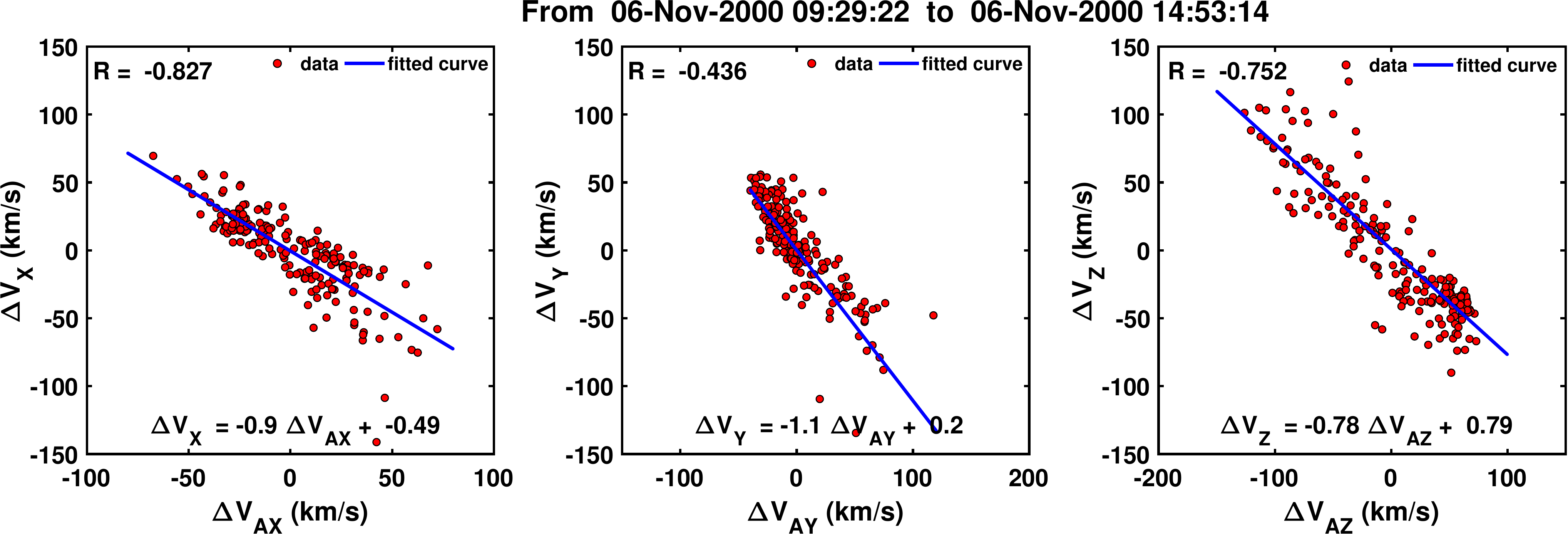}
	\caption{The correlation and regression analysis between the respective $\Delta V$ and $\Delta V_{A}$ components. The scattered black circle with filled red color represent observations from Wind spacecraft with time cadence of $92~s$. The $R$ is the  coefficient of correlation. The equation in each panel indicate the linear fit relation between respective components of $\Delta V$ and $\Delta V_{A}$.}
	\label{fig:Walen_p}
\end{figure}

In the first method, for the shock sheath region, the average values are estimated for each component of the magnetic field and solar wind vector respectively. We obtain $\Delta B$ by subtracting a mean value of the corresponding B component from each measurement. As a result, the Alfv\'en velocity fluctuation is given as: 
\begin{equation}
\Delta V_A = \frac{\Delta B}{\sqrt{\mu_0 \rho}}
\end{equation}
Similarly, we calculate $\Delta V$ by subtracting averaged proton flow velocity from measured values of each component respectively. In Figure \ref{fig:cpre} top three panels represents comparisons of $x$, $y$, and $z$ of components of $\Delta V_A$ and $\Delta V$ respectively. It clearly shows correlated variations between their respective components within the shock-sheath region and indicates the possibility of an Alfv\'en wave. Their correlation and regression analysis is depicted in Figure \ref{fig:Walen_p}. The Pearson correlation coefficients (R) of each x, y, and z components are –0.83,–0.44, and –0.75 and the corresponding regression slopes are –0.90, –1.1, and –0.78. (For ACE spacecraft data with 64 sec
time resolution, the correlation coefficients are –0.80, –0.92 and –0.91 and the corresponding regression slopes are –0.50,
–0.72, and –0.68 respectively. It should be noted that the corresponding figures are not displayed here. The Anti-Sunward Alfv\'en wave is confirmed by the significant negative correlation and regression coefficient  $\approx~-1$ confirms the presence of Anti-Sunward Alfv\'{e}n wave in the shock-sheath region of the examined CME \citep{gosling2010torsional,raghav2018first}. 
\subsection{Method 2}
In general, the mean values for selected regions or the average value of de Hoffmann-Teller (HT) frame are utilized as background quantities    \citep{raghav2018first,raghav2018torsional,raghav2018does,yang2013alfven,gosling2010torsional}. However, \citet{gosling2009one} and \citet{li2016new} suggested that the HT frame can change in high-speed solar wind streams and the solar wind fluctuations are pertinent to a slow varying base value of the magnetic field. In order to reduce the uncertainty in Alfvén wave identification, the fourth-order Butterworth bandpass-filters are applied to each component of plasma velocity and magnetic field data. The equally spaced 10 logarithmic frequency bands are selected. The applied filters are 10s-15s, 15s-25s, 25s-40s, 40s-60s, 60s-100s, 100s-160s, 160s-250s, 250s-400s, 400s-630s, and 630s-1000s. The  Wal\'en relation for each band-passed signal is analyzed as follows:
\begin{equation}
V_i = \pm R_w V_{Ai}
\end{equation}

\begin{figure}
	\includegraphics[width=\textwidth]{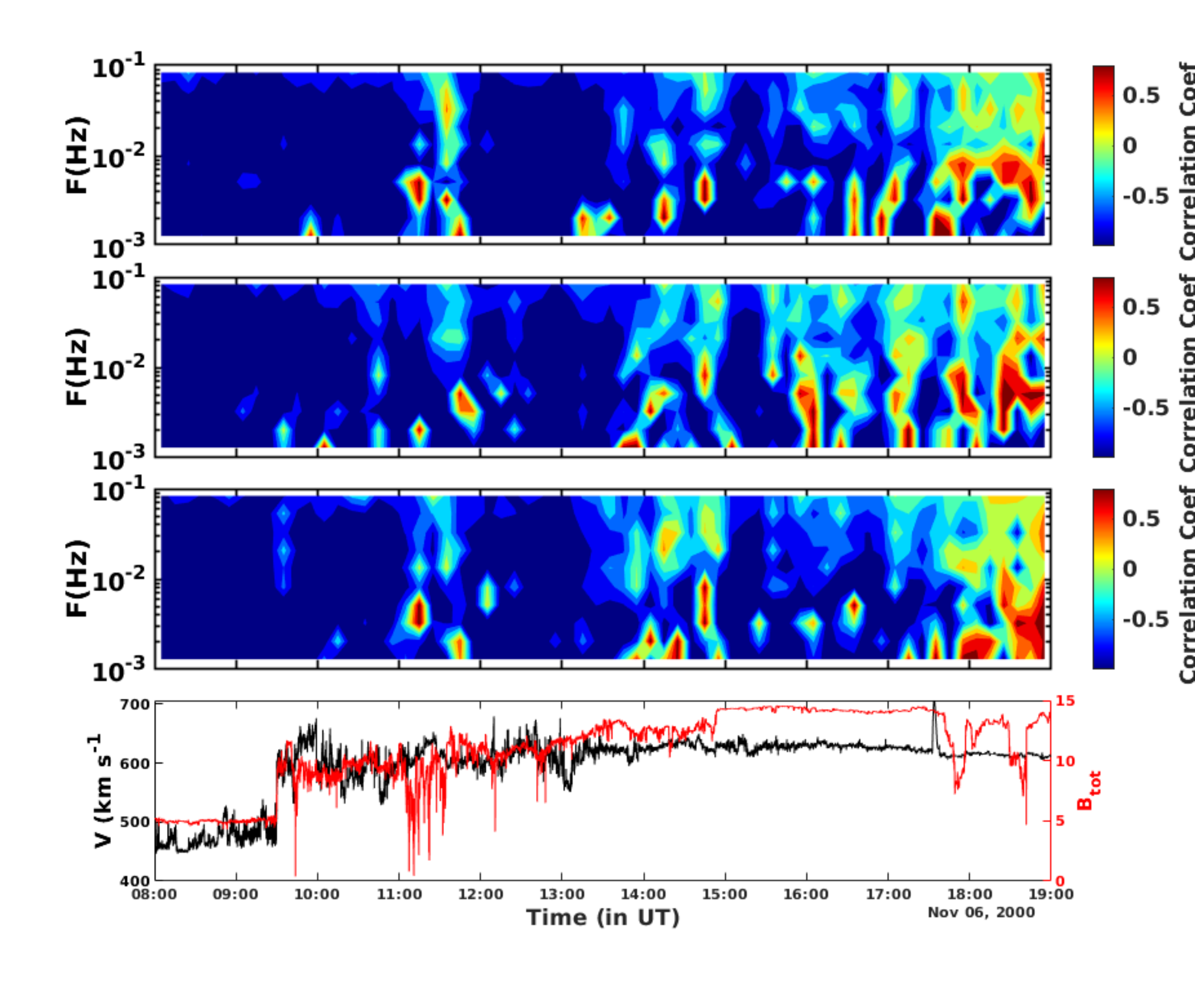}
	\caption{\textit{Time-frequency distribution of correlation coefficient between $ V_{Ai}$ and $V_i$ for complete event. Solar wind speed and magnetic field are shown in last panel. Wind satellite $3 s$  observations are utilized for the analysis. }}
	\label{fig:walen}
\end{figure}

The band-passed $V$ and $V_A$ components with the $i^{th}$ filter are represented here by $V_i$ and $V_{Ai}$. The value of the correlation coefficient between the respective components of $V_i$ and $V_{Ai}$ for each frequency band-passed signal confirms the presence of Alfv\'en waves or Alfv\'enic oscillations in the region under study. A similar approach is used by \citet{li2016new}. The complete region under study is separated with 15 minutes of time window into the sub-regions. The frequency-time distribution contour map with a 15-minute bin size is displayed in Figure ~\ref{fig:walen}. The value of the correlation coefficient for each sub-region is shown as a colour map on the contour plot. Dark blue shed shows negative correlation, while dark red shed shows a significant positive correlation. As a result, the presence of dominating Alfv\'enic flow along the x, y, and z components are shown in  Figure ~\ref{fig:walen} contour map by the dark color-strips, particularly inside the sheath region. 

\begin{figure}
\centering
		\includegraphics[width= 0.85\textwidth, height = 0.31\textheight]{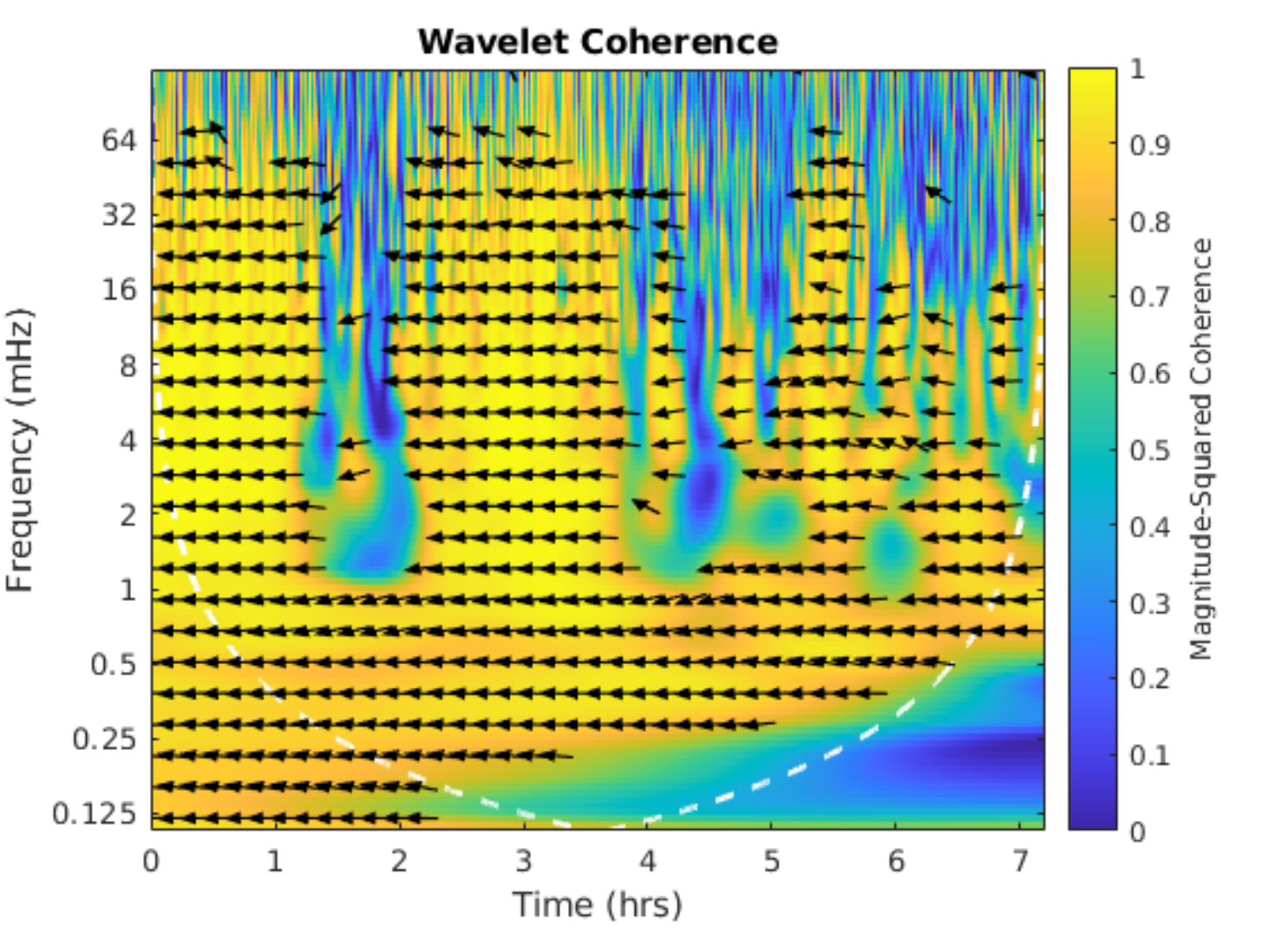}\\
	\includegraphics[width=0.85\textwidth, height = 0.31\textheight]{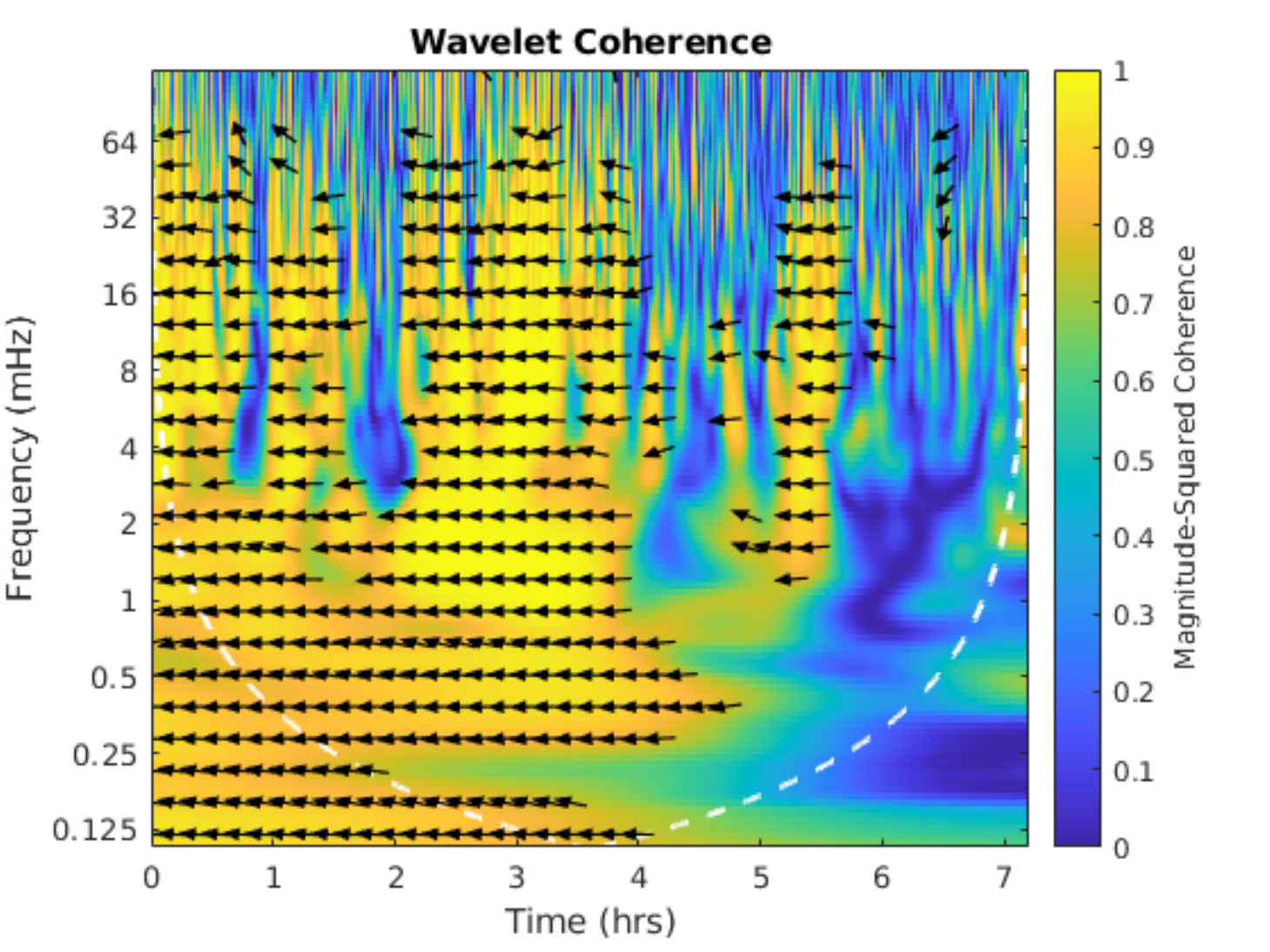}\\
	\includegraphics[width= 0.85\textwidth, height = 0.31\textheight]{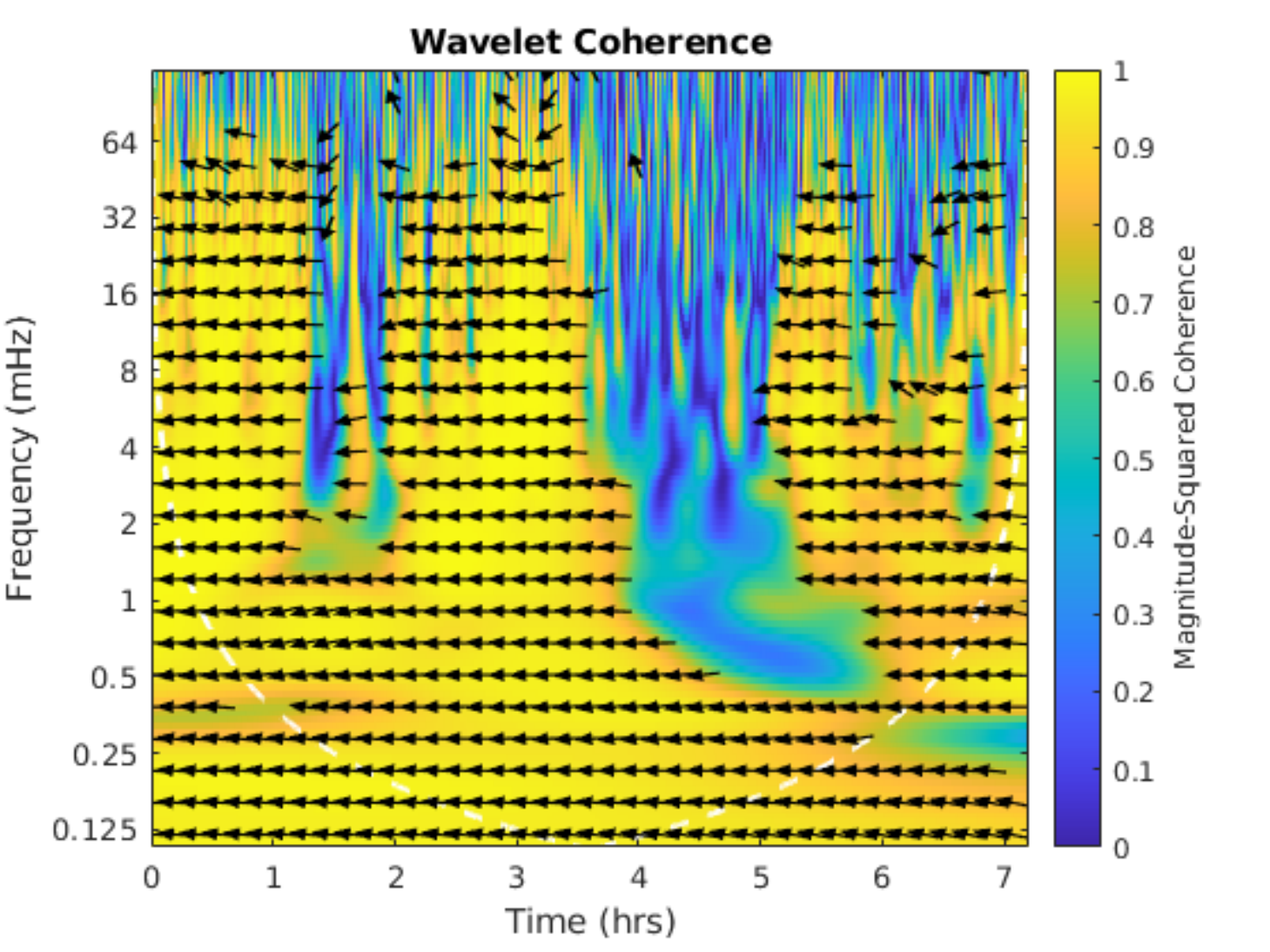}
	\caption{\textit{Wavelet coherence analysis between solar wind velocity and magnetic field components during ICME sheath and MC transit. The color bar represents  magnitude-squared coherence and angle of black arrows from the x + direction gives phase $\psi$ angle between the two signals.  The cone of influence is represented by the white dashed line marks. The onset of shock is treated as $0~ hours$, the front edge of MC is $\approx~ 5.5~ hours$}}
	\label{fig:waveletcoherence}
\end{figure}

 Very recently, \citep{chen2017nature} utilized a wavelet coherence test to study the nature of plasma turbulence at kinetic Alfv\'en scales in the Earth’s magnetosheath. The wavelet coherence test is generally employed to identify regions in time-frequency space where the two-time series co-vary (but does not necessarily have high power). We have also used this test to double-check the existence of Alfv\'enic fluctuations/waves in the ICME sheath region. Figure \ref{fig:waveletcoherence} demonstrate the magnitude squared wavelet coherence ($\gamma$), between $ V_{Ax}$ \& $V_x$ (top panel), $ V_{Ay}$ \& $V_y$ (middle panel), and $ V_{Az}$ \& $V_z$ (bottom panel) and the phase lag $\psi$ (black arrow) for $\gamma > 0.75$ (i.e. for highly significant correlated values), measured by Wind spacecraft. We observed a strong anti-correlation in all the components of $ V_{A}$ \& $V$ within the spacecraft frame frequencies range $0.25~mHz<f_{sc}<64~mHz$. This significant anti-correlation (see the direction of the black arrow in Figure ~\ref{fig:waveletcoherence}) clearly indicates that the shock-sheath region is dominated by the Anti-sunward deviation from anti-correlation is also visible for some frequency regions, especially towards the higher frequency side. Kindly note that for a certain interval of time (see middle and trailing edge of the sheath region), we observed the absence of Alfve\'nic fluctuations for certain frequency bands (for consistency, see Figure \ref{fig:walen} and \ref{fig:waveletcoherence}).

\section{Properties of Alfv\'en wave}

\begin{figure}
	\includegraphics[width=\textwidth]{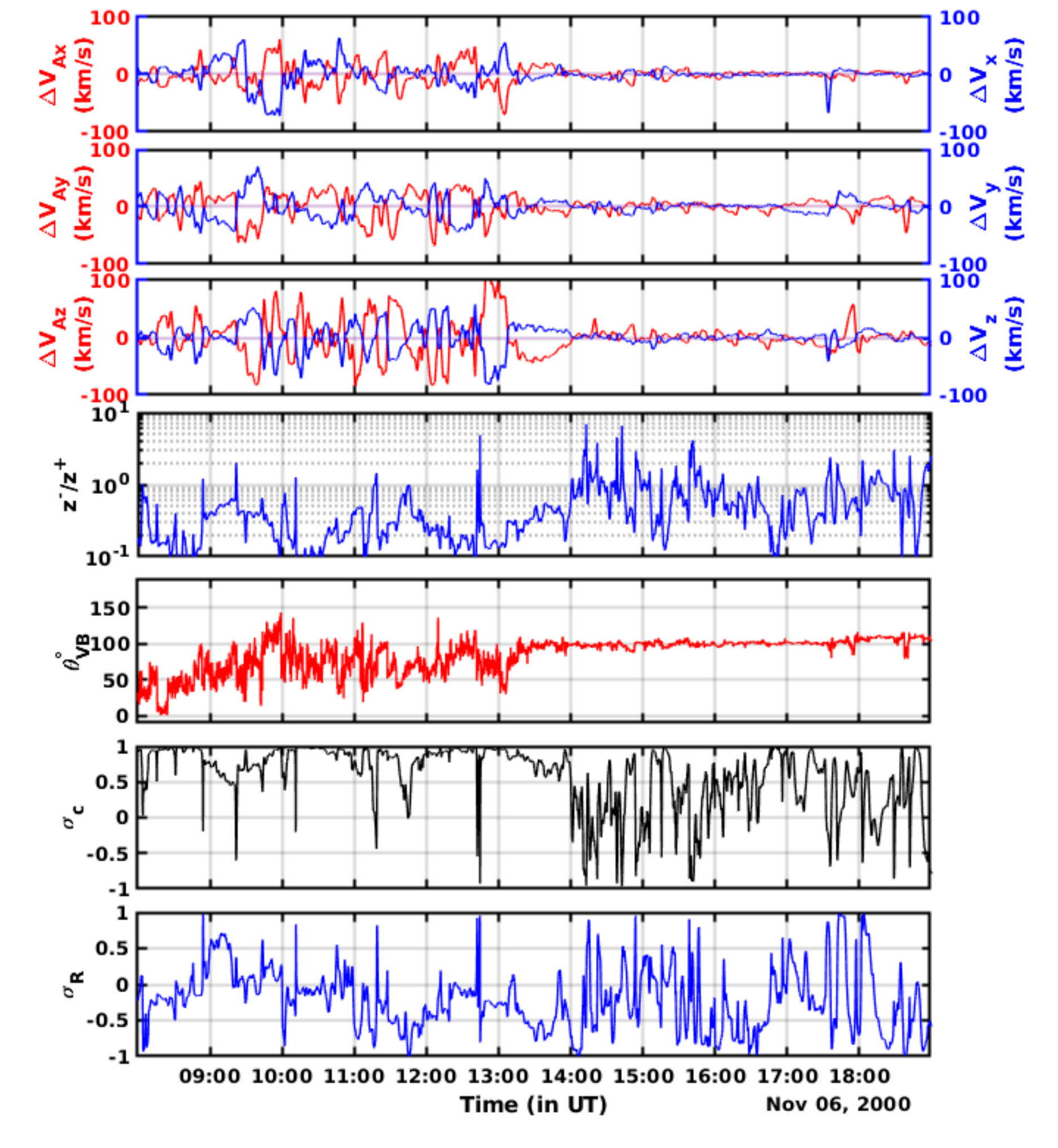}
	\caption{Top three panels demonstrate the temporal variation of Alfv\'{e}n velocity fluctuation vector $\Delta V_A$ (red)  and that of the proton flow velocity fluctuation vector $\Delta V$ (blue). It demonstrates  Alfv\'{e}nic and shock-sheath characteristic of the studied region of an ICME. The fourth panel represents the  ratio of inward to outward  El\"{a}sser variable. The fifth panel gives appearance of the angle between solar wind velocity and magnetic field. The Sixth panel and seventh  panels represents the temporal variation of the normalized cross helicity ($\sigma_c$) and normalized residual energy ($\sigma_R$) respectively. The analysis is performed using wind spacecraft data with time cadence of $3~sec$.} 
	\label{fig:cpre}
\end{figure}

Here, we opine that the Wal\'{e}n test and definition of Alfv\'{e}nicity are based to a large extent, on the approximate incompressibility of the background. The top three panels of figure  \ref{fig:cpre} make it very evident how closely correlated  the components of $\Delta V$ and $\Delta V_A$ are to one another. 

Presence of  Alfv\'enic fluctuation within the solar wind (especially in corotating high velocity streams) demands to use Els$\ddot{a}$sser variables to separate out the  contribution of ``outward'' and ``inward'' to the turbulence. Els$\ddot{a}$sser variables are used in the theoretical studies \citep{elsasser1950hydromagnetic,dobrowolny1980fully,dobrowolny1980properties,veltri1982cross,marsch1987ideal,zhou1989non} as well as for the first time in interplanetary space (data analysis) by \citet{grappin1990origin,tu1989basic,tu1990evidence}. The Els$\ddot{a}$sser variables are defined as: 
\begin{equation} \label{zpm}
\vec{Z}^{\pm} = \vec{V} \pm \frac{\vec{B}}{\sqrt{4\pi \rho}}, 
\end{equation}

here, $\vec{V}$  and $\vec{B}$ are proton velocity and magnetic field fluctuation, measured in the GSE co-ordinate system. The $\pm$ sign in front of $\vec{B}$ depends on the sign of [$-k ~\cdotp B_0$]. The Els$\ddot{a}$sser variables are defined in such a way that $\vec{Z}^+$ and $\vec{Z}^-$ always refers to the waves propagating outward and inward direction \citep{roberts1987origin,roberts1987nature}.  $\theta_{VB}$ is estimated as
\begin{equation}
	\theta_{VB}=cos^{-1}(\frac{-B_x}{B_{mag}})
\end{equation}
 The equation \ref{zpm} get modified for $\theta_{VB} \leq 90^\circ$ as $\vec{z^{+}}=\vec V -\vec V_A$ and $\vec{z^{-}}=\vec V +\vec V_A$ , for $\theta_{VB} > 90^\circ$ the equation will remains the same.  
%
The energy associated with $z^+$ and $z^-$ is defined as: 
\begin{equation}
e^\pm = \frac{1}{2} \langle (z^\pm)^2 \rangle,
\end{equation} 
Moreover, the normalized cross helicity is estimated as 
\begin{equation}
\sigma_c = \frac{e^+ - e^-}{e^+ + e^- },
\end{equation}
also, the normalized residual energy defined as 
\begin{equation}
	\sigma_R=\frac{e^v-e^b}{e^v+e^b}
\end{equation} 
where $e^v$ $\&$ $e^b$ is kinetic and magnetic energy respectively. $\sigma_R$ is measure of the excess magnetic field energy with respect to kinetic energy or vice versa \cite{bruno2013solar}. Where, $e^v = 0.5 < v^2>$ and $e^b = 0.5 < b^2>$ are  kinetic and magnetic energy associated with Els\"{a}sser variables ($z^\pm$). Here, $v = 0.5(z^+ + z^-)$ and $b = 0.5(z^+ - z^-)$ respectively.

Here, the $\sigma_c > 0$ $\&$ $\sigma_R < 0$ indicates dominant flow of outward propagating waves \citep{matthaeus1982measurement,tu1989basic}. In our study, we got highly positive value  of normalized cross helicity in the shock-sheath region indicates dominance of outward Alfv\'enic turbulence.

\begin{figure}
	\includegraphics[width=\textwidth]{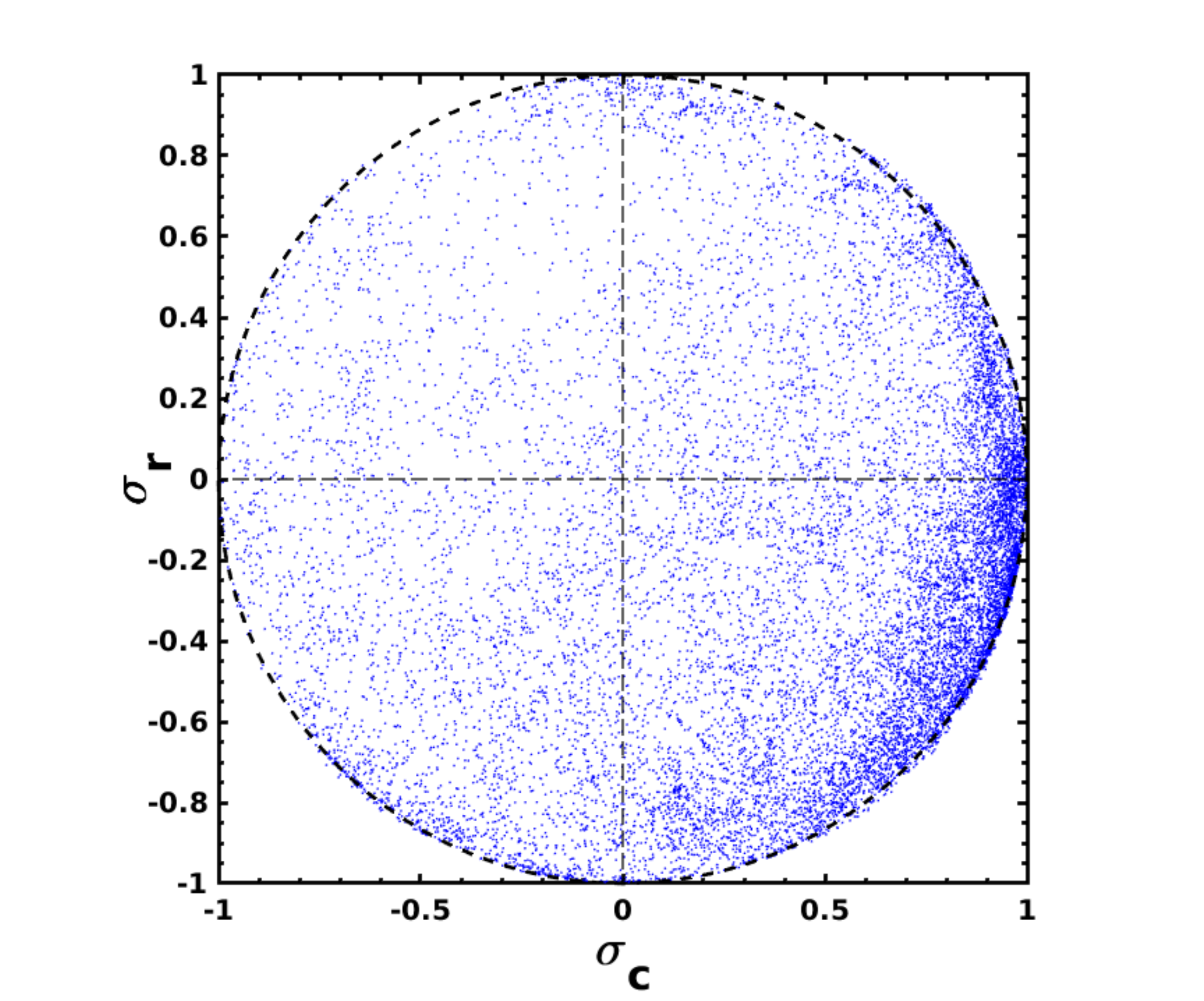}
	\caption{The scatter plot of normalized residual energy ($\sigma_r$) and normalized cross helicity ($\sigma_c$) , estimated by Wind spacecraft data  of 2000 November 06.}
	\label{fig:sigma_plot}
\end{figure}

The Figure \ref{fig:sigma_plot} demonstrate scattered plot of the normalized residual energy ($\sigma_R$) Vs normalized cross helicity ($\sigma_c$). These defined parameters are valid only if $\sigma_R^2 + \sigma_c^2 \leq 1$ \textit{i.e.}, the data points should lie within the circle \citep{bavassano1998cross}. Note that most of the  observed data points has positive $\sigma_c$. This implies that identified Alfv\'en wave within the ICME sheath region dominantly propagating towards the earth. 
\section{PSD Analysis}

\begin{figure}
	\includegraphics[width=\textwidth]{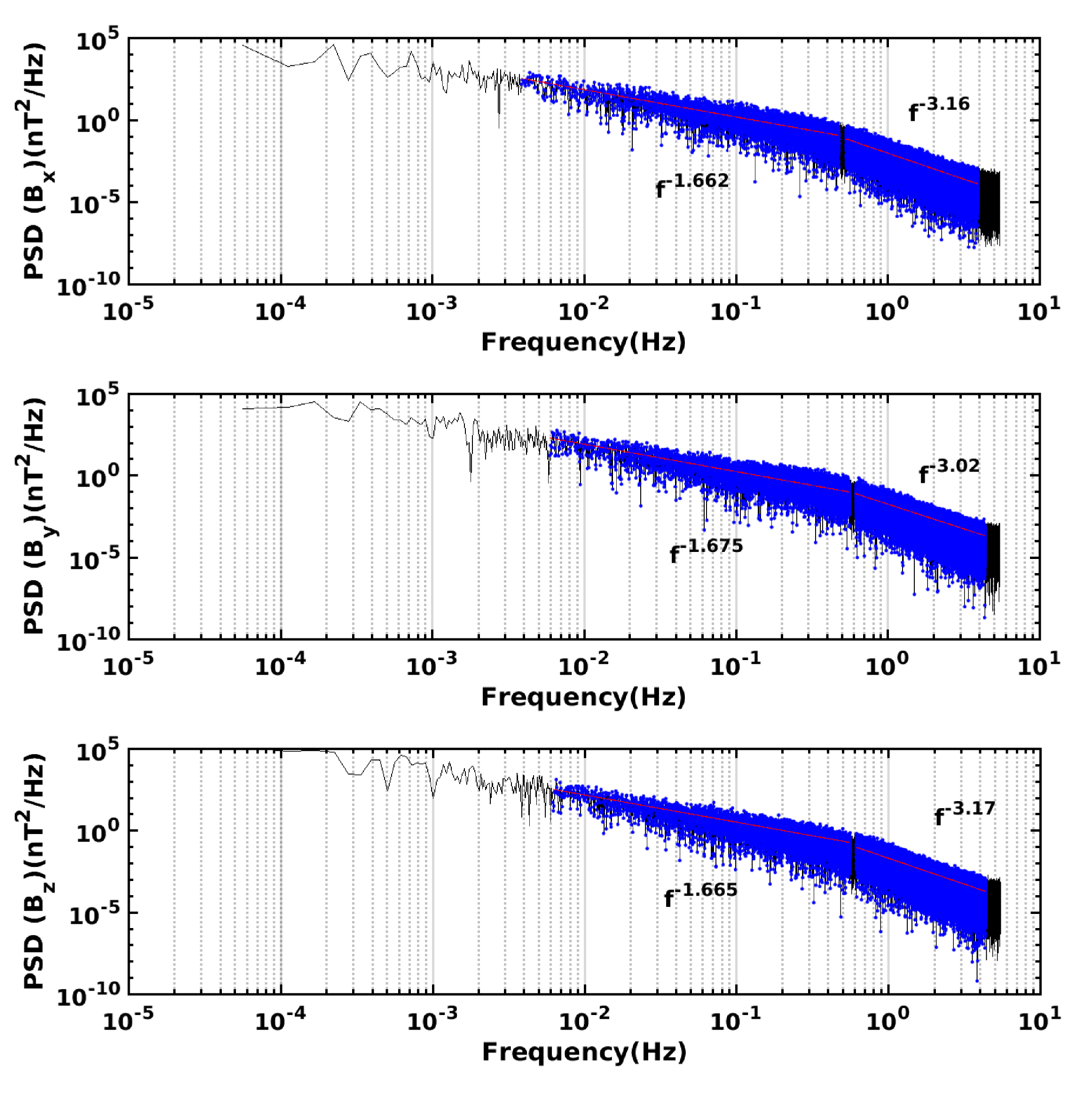}
	\caption{Power spectra of magnetic fluctuations along the $B_x$ direction in GSE coordinate (black color) as a function of frequency in the spacecraft frame as measured by Wind on 2000 November 06, from 09:15 to 14:16 UTC, computed with FFT (black) algorithms.  We have used WINDS high resolution (11 Hz) IMF data.}
	\label{fig:PSD}
\end{figure}

We performed a power spectral density (PSD) analysis to study the characterization of the multi-scale nature of shock-sheath turbulences/fluctuations. Figure \ref{fig:PSD} represents the spectral output for the $B_x$, $B_y$ and $B_z$ components of the IMF. The common feature of the incompressible MHD plasma turbulence i.e. Kolmogorov-like turbulence is observed in the intermediate frequency range which is consistent with $\sim f^{-1.66}$ \citep{bruno2005solar}. We believed that the entire turbulent interactions within these regimes are governed by the Alfv\'{e}nic cascade. 

We estimate cyclotron frequencies for proton and alpha particle for the observed magnetic field of $5-12~nT$ range, which turn out to be $0.47 ~Hz - 1.10~Hz$  for proton and $0.24 ~Hz - 0.55~Hz$ for alpha. The various studies reveal that at length-scales beyond the  MHD regime, the power spectrum shows spectral break which halts the Alfv\'{e}nic cascade \citep{goldstein1994properties,leamon1999dissipation,bale2005measurement,alexandrova2008small,sahraoui2009evidence,shaikh20093d}. At $\sim 0.5 -0.6$ Hz, the PSD spectrum ``breaks" from a $\sim f^{-1.6}$ power-law inertial range to a $\sim f^{-3.1}$  dissipation range (see Figure \ref{fig:PSD}). However, \cite{perri2010does} suggest that the spectral break in the solar wind is independent of the distance from the Sun and that of both the ion-cyclotron frequency and the proton gyro-radius. Therefore, it is also possible that the observed high-frequency break in our study is caused by a combination of different physical processes as a result of high compression within the shock-sheath region. The other possible mechanism for the spectral break may result from energy transfer processes related to; 1) kinetic Alfv\'{e}n wave (KAW) \citep{hasegawa1976parametric}, 2) electromagnetic ion-cyclotron-Alfv\'{e}n (EMICA) waves \citep{wu2007proton,gary2008cascade}, or the fluctuation associated with the  Hall  magnetohydrodynamics (HMHD) plasma model \citep{alexandrova2007solar,alexandrova2008small}.

At higher frequencies, the spectrum of the magnetic field fluctuations has power-law dependence as $\sim f^{-\alpha}$, where, the value of $\alpha$ may range from 2 to 4. The average value of the $\alpha $ is close to 7/3 \citep{leamon1998observational,smith2006dependence,alexandrova2008small}. In our study, it is about $\sim {-3.1}$. These higher frequency part of the spectrum may be associated either with a dissipative range \citep{leamon1998observational,smith2006dependence} or with a different turbulent energy cascade caused by dispersive effects \citep{stasiewicz2000small,sahraoui2006anisotropic,galtier2007multiscale,alexandrova2008small,sahraoui2009evidence,li2016new}. \citet{stawicki2001solar}  proposed that suppression of the Alfv\'{e}nic fluctuations are due to the ion cyclotron damping at intermediate wave frequency (wavenumber), hence the observed power spectra are weakly damped dispersive magnetosonic and/or whistler waves (unlike Alfv\'en waves). The presence of the whistler wave mode in the high-frequency regime was proposed by the \citep{beinroth1981properties}. \citet{goldstein1994properties} found out the existence of multi-scale waves (Alfv\'enic, whistlers, and cyclotron waves) with a single polarization in the dissipation regime of the spectrum. Observation of the obliquely propagating KAWs (in the $\omega < \omega_{ci}$ regime or Alfv\'enic regime) puts a question about the spectral breakpoint due to damping of ion cyclotron waves \citep{howes2008kinetic}. The Kinetic \citep{howes2008kinetic} and Fluid  \citep{shaikh20093d} simulations show that the ion inertial length-scale is comparable to that of the spectral breakpoint near the characteristic turbulent length-scales. For the length-scales larger than the ion inertial length-scales, the simulations demonstrate  Kolmogorov-like spectra. Moreover, for smaller ion inertial length-scales, they observed the steeper spectrum that is close to $f^{-7/3}$.

\section{Discussion }

Using three distinct techniques, we found  a substantial correlation between the variations in the magnetic field and velocity vectors. This suggests that the magnetic field and fluid velocity are oscillating simultaneously and propagating in the same direction as the magnetic tension force. The various plasma features (see Figure ~\ref{fig:Walen_p}, \ref{fig:walen},\ref{fig:waveletcoherence},\ref{fig:cpre}) confirms the presence of  Alfv\'{e}n wave in shock-sheath region. Similar analysis has been performed for various ICME-sheath regions observed by Wind spacecraft and listed in \url{https://wind.nasa.gov/ICMEindex.php}. It is important to note that Alfv\'en waves/fluctuations have not been found in all the studied ICME-sheath events. Moreover, only following events shows inward or outward flow of Alfv\'en waves/fluctuations out of studied events; e.g Very soon we will analysed all the events listed in aforementioned catalogue to study their possible origin, propagation and dissipation in sheath plasma.


The overall distinguishable features of Alfv\'en waves during the shock sheath strongly support dominant Alfv\'enic turbulence. Moreover, the Alfv\'{e}n waves are pervasive in the solar wind, and it is important to note that the method 2 (section 3.2, Figure ~\ref{fig:walen}) shows the presence of Alfv\'en wave in up-stream of shock. However, their transmission in shock-sheath is questionable. 
The solar wind at 1 AU is overwhelmingly Alfv\'enic, therefore it is possible that the same Alfv\'enic background is present in the ICME shock-sheath region too and a     significant compressible component is just superposed on that \citep{chen2013residual}. 

The  Alfv\'{e}n waves lead to non-linear interactions \citep{dobrowolny1980fully} which are crucial for the dynamical evolution of a Kolmogorov-like MHD spectrum \citep{bruno2013solar}. We have also performed a power spectral density (PSD) analysis for all IMF vector components. It depicts Kolmogorov-like turbulent nature (The PSD analysis follows $f^{-1.66}$ spectrum) for
the frequency range between $0.4\times10^{-3}$ Hz to 0.5 Hz in the
studied the shock-sheath region. Thus the existence of Alfv\'en waves with the Kolmogorov-like turbulence depicts Alfv\'enic turbulent nature of the shock-sheath. Thus, we observed the continued cascade of energy from large scales to smaller scales of wavelengths and eventually to such small scales that the plasma no longer behaves like a fluid due to a change in velocity and magnetic field fluctuations. At this scale, the particle distribution is affected by the magnetic field which may lead to plasma heating through resonant interactions \citep{tsurutani1997some}. We opine that the plasma heating in shock-sheath could be associated with an above-discussed process. 

Alfv\'enic-like fluctuations may also occur in the impact sheath, but the dominant one will be the quasi-2D structure \citep{zank2018theory}. For example, \citet{zheng2018observational} used the Grad-Shafranov reconstruction method to show the flux rope in the impact sheath area. However, our observations clearly demonstrate the existence of Alfv\'en waves in the sheath region. It does not mean that Alfv\'enic fluctuations dominantly seen in the sheath, rather a statistical study will put some light on the aforementioned issue.

\section{Implications}
Several open questions need to be addressed in view of turbulent nature in highly compressed and heated shock-sheath such as, (i) What is the origin of a turbulent cascade in shock-sheath?  Is it the coronal plasma or local driving?; (ii) How does the cascade modify the shock-sheath plasma?; (iii) How do the turbulent fluctuations get dissipated into heat?; (iv) What is more important for energy dissipation, non-linear turbulent heating, or resonant wave-particle interactions?; (v) Can shock-sheath turbulence be parameterized and included in heliospheric models for space weather prediction? 

Recently, the presence of the Alfv\'{e}n wave has been seen in a magnetic cloud of CME \citep{raghav2018first}. It is manifested that the Alfv\'{e}nic oscillations in a magnetic cloud of CME may cause the internal magnetic reconnection and/or thermal anisotropy in plasma distribution which leads to the disruption of the stable magnetic structure of the CME \citep{raghav2018does}. Their presence in the magnetic cloud of CME also controls the recovery phase of the geomagnetic storms \citep{raghav2018torsional,raghav2019cause}. In the introduction section of the article, we emphasize that the shock-sheath of CME not only affects the interplanetary plasma characteristics but also affects the dynamics of the magnetosphere, ionosphere, radiation-belt, and upper atmosphere of the Earth. It affects the other planetary exospheres as well. Therefore how the typical configuration such as Alfv\'{e}n fluctuations embedded shock-sheath influence the overall solar-terrestrial plasma will be intriguing and may activate the possible direction of future studies. One can also expect similar features of shock-sheath in interstellar medium as well e.g. supernovae shocks and associated sheaths.\\






\ldots

\section{acknowledgement}
We acknowledge that the NASA/GSFC Space Physics Data Facility makes the OMNIWeb (or CDAWeb, or ftp) service and OMNI data available. We thank SERB, India for supporting AR and OD through project reference file number CRG/2020/002314. We also acknowledge the Department of Physics (Autonomous), University of Mumbai, for supplying the resources necessary to complete this research.

\bibliographystyle{unsrtnat}
\bibliography{shock}

\end{document}